\documentclass[a4paper]{jpconf}
\usepackage{graphicx}
\usepackage{wrapfig}

\begin{document}
\title{Transfer Learning in Astronomy: \\ A New Machine-Learning Paradigm}

\author{Ricardo Vilalta}

\address{
Department of Computer Science, University of Houston. \\
Philip G. Hoffman Hall, 3551 Cullen Blvd., Houston TX, 77204-3010, USA}

\ead{rvilalta@uh.edu}

\begin{abstract}
The widespread dissemination of machine learning tools in science, particularly in astronomy, has revealed the limitation of working with simple single-task scenarios in which any task in need of a predictive model is looked in isolation, and ignores the existence of other similar tasks. In contrast, a new generation of techniques is emerging where predictive models can take advantage of previous experience to leverage information from similar tasks. The new emerging area is referred to as \textit{transfer learning}.  In this paper, I briefly describe the motivation behind the use of transfer learning techniques, and explain how such techniques can be used to solve popular problems in astronomy. As an example, a prevalent problem in astronomy is to estimate the class of an object (e.g., Supernova Ia) using a generation of photometric light-curve datasets where data abounds, but class labels are scarce; such analysis can benefit from spectroscopic data where class labels are known with high confidence, but the data sample is small. Transfer learning provides a robust and practical solution to leverage information from one domain to improve the accuracy of a model built on a different domain. In the example above, transfer learning would look to overcome the difficulty in the compatibility of models between spectroscopic data and photometric data, since data properties such as size, class priors, and underlying distributions, are all expected to be significantly different.
\end{abstract}


\section{Introduction}
\label{sc:introduction}

The abundance of large datasets generated for scientific research through sophisticated sensors (e.g., modern telescopes) or complex simulations, has led to a widespread interest for automated mechanisms that can analyze the data and generate models that classify events in an accurate and precise manner. A popular approach uses machine learning \cite{Hastie09} to train a computer to recognize different events using case examples that belong to different categories. In most applications, a common assumption pervading most traditional work in machine learning is that the probability distribution from which a training sample is drawn is static; future samples must follow the same distribution for any model to remain valid. While such assumption is sensible, and has found a plethora of successful real-world applications, recent work in machine learning has shown an equally large number of applications that do not follow such assumption.

An example of such shift in distributions lies in light curve classification from star samples obtained from different galaxies \cite{Vilalta13}. An original source task may consist of identifying certain types of stars on a nearby galaxy. But if we move to galaxies lying farther away, and we try to repeat the same identification task, we will find that the distribution of stars has now changed (sometimes drastically). A major reason for such change is that at greater distances, less luminous stars fall below the detection threshold and more luminous stars are preferentially detected. The corresponding change in distribution precludes the direct utilization of one single model across galaxies; it calls for a form of model adaptation to compensate for the change in the data distribution. Many other domains exist where changes in the underlying probability distribution are primarily caused by measurements obtained under different circumstances (e.g., observing galaxies that lie at different distances), by changing the orientation or position of the same sensing device, or by utilizing a similar but more powerful device. In particle physics, for example, a model built to identify a certain particle is rendered unapplicable when we collect samples obtained using more powerful particle accelerators; this is because the range of parameter values shifts as we reach out to higher energies.

The abundance of examples exhibiting a shift in distribution as circumstances change over time calls for automated methods that adapt as a response to the dynamic nature of many learning tasks. The type of problems mentioned above have recently led to the development of a new area of study called \textit{transfer learning}, where the mechanism leverages previous experience to increase the accuracy of predictive models. In this paper, I briefly introduce central ideas in transfer learning and describe a practical application in astronomy that captures the essence of this new paradigm.

\begin{figure}[bt]
\vspace*{-3mm}
\hspace*{5mm}
\includegraphics[scale=0.25]{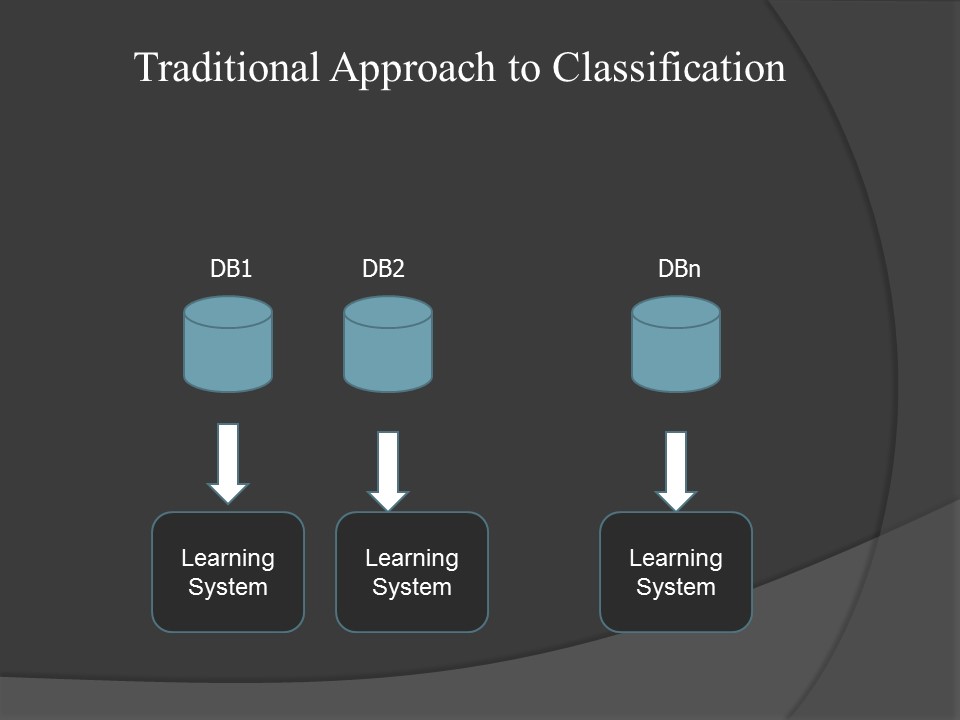}
\hspace*{15mm}
\includegraphics[scale=0.25]{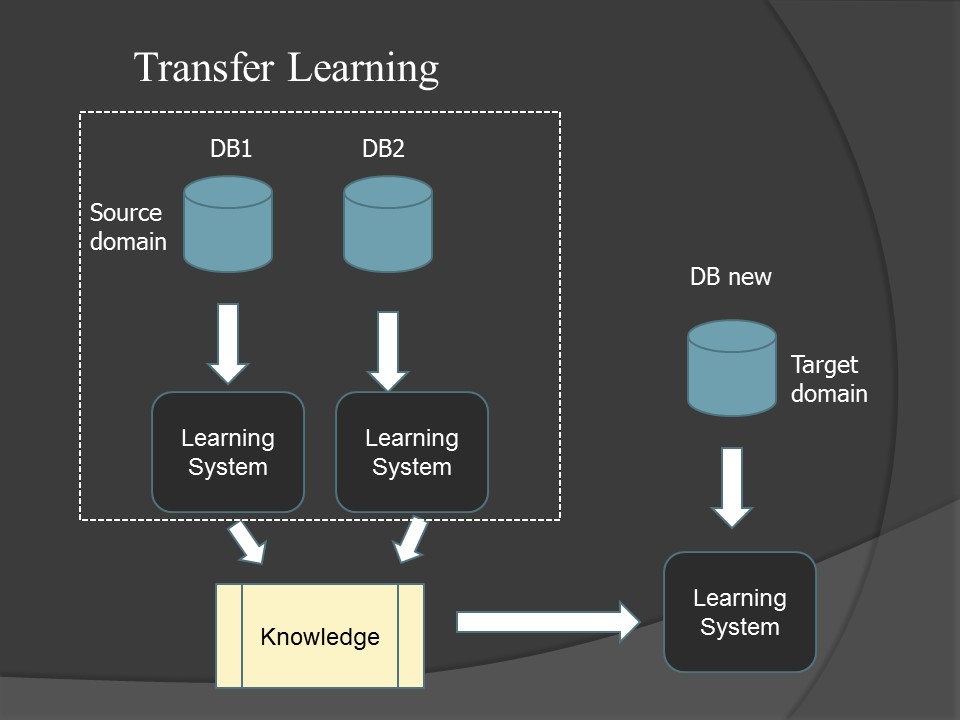}
\caption{\small (left) Traditional machine learning considers each task in isolation. (right) Transfer learning leverages
information from previous experience.}
\vspace*{-2mm}
\label{fig:TransferLearning}
\end{figure}

\subsection{Rationale for Transfer Learning}

Before the advent of transfer learning techniques, practical solutions to the problem of shifts in probability distributions led to a continuous re-training of a predictive model over time. Each new instance of a classification or regression problem was considered in isolation, lacking the capacity to exploit information from previous experience. Figure~\ref{fig:TransferLearning}(left) shows this operational mode. A training set serves as input to the process of modelling data using machine learning tools. Other datasets are treated similarly but in isolation. No information is exploited that can serve to learn experience across tasks. The traditional approach is particularly detrimental in situations where similar datasets share information that is useful across all tasks. This is clearly the case in the astronomical or particle-physics scenarios mentioned above, where a change in the sensor device or the event generator does not render any previous form of data analysis useless; instead one should be able to exploit the presence of similar patterns across tasks.

The new approach to the construction of adaptive learning models is to gather experience from previous tasks to improve on the current target task. This is also known as the transfer learning problem. Figure~\ref{fig:TransferLearning}(right) depicts the new approach. Previous experience is now retained for future analysis (source domains). The repetitive use of machine learning is stored in a knowledge database that varies in nature according to the problem or task under analysis. The main difference with the previous approach becomes evident when a new target domain arrives (i.e., new task). The data modelling process is now reinforced with previous experience to strengthen the predictive model, and gain leverage from patterns found in previous tasks. A brief description of these adaptive techniques is given in the following section.


\section{Different Modalities in Transfer Learning}
\label{sc:types-tl}

Before describing different approaches to transfer learning, I introduce some notation. It is common to assume a source dataset $T_s = \{(\mathbf{x}, y)\}$ made of training examples, where $\mathbf{x} \in \mathcal{X}$ is a feature vector, and $y \in \mathcal{Y}$ is a class or category. Dataset $T_s$, the source dataset,  corresponds to a previous application of machine learning that produced a predictive model $f_s(\mathbf{x})$. Examples in $T_s$ are drawn randomly from a joint distribution $P_s(\mathbf{x},y)$; of relevance here are the marginal distribution $P_s(\mathbf{x})$ and the class-posterior distribution $P_s(y|\mathbf{x})$. We will assume model $f_s(\mathbf{x})$ needs some form of adaptation because of changes in the original data distribution. The second dataset, $T_t$, the target dataset, corresponds to a new application that is similar to the source task, but not identical. We assume the existence of corresponding distributions on the target task: marginal distribution $P_t(\mathbf{x})$, and class-posterior distribution $P_t(y|\mathbf{x})$. Now, rather than building $f_t(\mathbf{x})$ from scratch, transfer learning can be invoked to exploit previous experience \cite{BenDavid10,Blitzer08}. A justification for the use of transfer learning is that $T_t$ is either small, or contains few --or none-- labeled instances (e.g., due to the cost of label annotation).

Many techniques have been proposed in the area of transfer learning \cite{Pan10,Weiss16}. A broad taxonomy begins by defining two approaches: homogenous and heterogenous transfer learning. In homogenous transfer learning, the feature representation for source and target is identical $\mathcal{X}_s = \mathcal{X}_t$ (i.e., both tasks share the same feature representation). In heterogenous transfer learning we dispense with such assumption.

Within homogenous transfer learning, a simple scenario assumes the difference between source and target lies on the marginal distributions $P_s(\mathbf{x}) \neq P_t(\mathbf{x})$. Other scenarios assume the marginals remain the same, while the class-posterior distributions differ $P_s(y|\mathbf{x}) \neq P_t(y|\mathbf{x})$, or that both marginals and posterior distributions differ.

A final categorization of transfer learning methods focuses on the nature of the element being transferred. One approach is to transfer instances between source and target. The strategy behind instance-based methods is to increase the weight of source instances populating regions of high density in the target domain. A piece of work along these lines is known as covariate shift \cite{Quinonero09,Shimodaira00}. Under the covariance-shift assumption, the model built on the new weighted source distribution can be directly applied to the target domain. A strong assumption here requires close proximity between source and target distributions.

Another approach is to transfer features across tasks. Feature-based transfer learning projects both the source and the target datasets into a common feature space where the covariate-shift assumption holds. The new model built on the transformed space acts as the classifier on the target. Instances of this family include subspace alignment methods, where the goal is to find a common subspace that makes source and target distributions overlap.

A third approach is to transfer parameter values across tasks. This can be accomplished by simply transferring parameter values directly from source to target, or by following a Bayesian approach where model parameters on the source task are used to build a prior distribution on the target task. Additionally, multiple models can be induced from the source task, and a weighting scheme can be used to provide adequate weights to combine the learners on the target task.

The last approach is to transfer knowledge based on the nature of the relation between tasks. As an example, in relational-based transfer learning, the idea is to look for relational patterns across source and target. An example is to predict documents based on grammatical and sentence structure patterns across texts that belong to similar topics \cite{Li12}.


\newpage
\section{An Application of Transfer Learning in Astronomy}
\label{sc:application-tl}

An application of transfer learning in astronomy can be found in the automated identification of Supernova Ia (SNe~Ia) \cite{Gupta16}. The task is of great importance to astronomy, because SNe~Ia are considered standard candles in probing large cosmological distances; the correlation between their luminosity and distance independent quantities has led to multiple discoveries, including the accelerating expansion of the Universe \cite{Riess98}. The discrepancy in distribution that makes this a problem amenable to transfer learning is due to the difference between spectroscopic and photometric observations. Spectroscopy provides a high-resolution description of electromagnetic radiation and is crucial to estimate chemical composition (through spectral lines) and distances (redshift) with high precision. But spectroscopic observations are costly and time-consuming; in practice, it is more convenient to perform photometric observations that summarize radiation in a set of broad wavelength windows or filters, even at the expense of forfeiting the wealth of information otherwise available using a spectroscopic analysis.

Many modern surveys including the the Sloan Digital Sky Survey, the Dark Energy Survey, and the upcoming Large Synoptic Survey (Telescope) are designed to capture  pure photometric observations for SNe Ia. A big challenge in modern astronomy is to infer spectroscopic properties from purely photometric data. In our current discussion, the source domain corresponds to spectroscopic data where Supernovae are confidently classified (as type Ia or different); the target domain corresponds to the new generation of abundant photometric light-curve datasets where class labels are scarce.  Our goal is to take advantage of the source domain to attain high predictive performance on the target domain. The use of transfer learning here is crucial, since photometric observations lack a precise class label (e.g. type of SNe) to conform a reliable training set. At the same time, the  naive approach of training a model $f_s(\mathbf{x})$ on the spectroscopic data $T_s$ (source domain) and applying it directly on the photometric data $T_t$ (target domain) is prone to failure because of the distributional shift observed between both types of observations.

\subsection*{Methodology}

I now describe a specific study aimed at building accurate SNe Ia classifiers from photometric data, while exploiting information from spectroscopic data \cite{Gupta16}. The first step is (almost always invariably) to reduce the dimensionality of the (source and target) data, since the original samples contain a large number of features. A recommended approach is to use kernel principal component analysis (Kernel PCA) \cite{Scholkopf97}; the technique is a generalization of principal component analysis (PCA) by using non-linear components. Recently, new approaches have emerged that compute robust non-linear combinations of features, such as those found in deep-learning architectures \cite{Goodfellow16,Sasdelli16}.

The second step is to use transfer learning to leverage experience from the source domain (spectroscopic dataset) to attain an accurate classifier on the target domain (photometric dataset). The study reported here considered two methods used: kernel mean matching KMM, and subspace alignment SA. The first method, KMM, works through a re-weighting scheme that gives more importance to those examples on the source dataset that appear closer to the target dataset \cite{Gretton09}. Specifically, the method projects the data into a new space (reproducing Hilbert kernel space) where it minimizes the (maximum) distance between the means on each distribution (source and target). The projection helps to identify source examples that can be incorporated into the training phase while building a classifier on the target domain.

The second method, subspace alignment, is a common transfer learning method based on feature subset selection \cite{Fernando14}.
The idea is to apply PCA on source $T_s$ and target $T_t$ datasets separately by choosing a common space. It then attempts to align the projected  source dataset with the projected target dataset in this common  subspace using a subspace alignment matrix. Once source and target are aligned, a classifier is built on the transformed source dataset $T_s^{\alpha}$, and subsequently applied to the transformed target dataset $T_t^{\alpha}$.

\subsection*{Empirical Results}

\begin{wrapfigure}{r}{0.5\textwidth}
\vspace*{-4mm}
\includegraphics[scale=0.20]{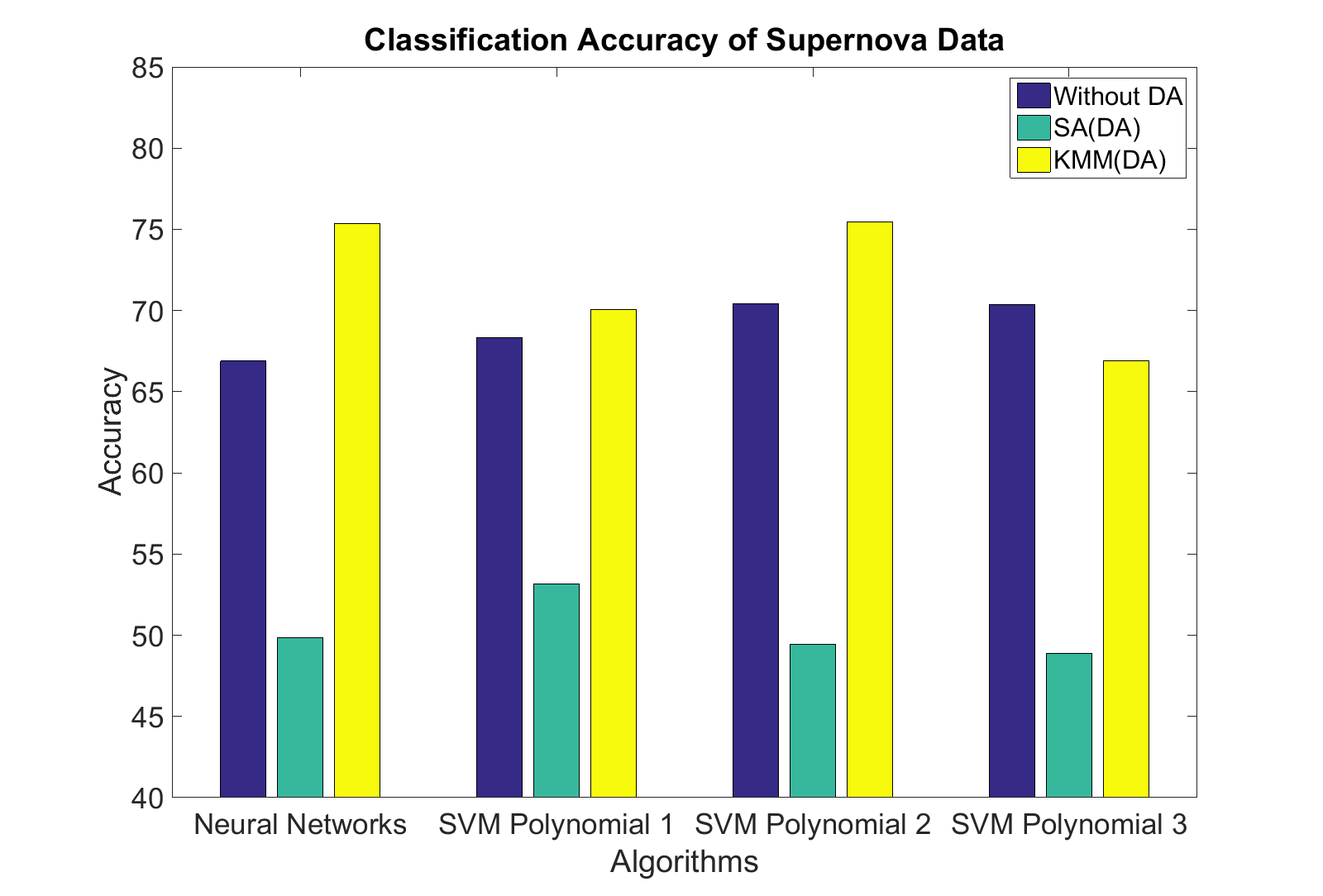}
\vspace*{-5mm}
\caption{\small Accuracy on supernova target dataset with and without transfer learning (labeled as DA for Domain Adaptation).}
\label{fig:SNeAccuracy}
\vspace*{-4mm}
\end{wrapfigure}

Experimental results are shown in Figure~\ref{fig:SNeAccuracy}. The horizontal axis shows different classifiers including neural networks and support vector machines \cite{Hastie09}, the latter implemented with different degrees of complexity.  The vertical axis shows accuracy (proportion of examples correctly classified) using 10-fold cross validation.  Some interesting observations in Figure~\ref{fig:SNeAccuracy} are as follows. First, KMM does provide a significant advantage over a methodology that lacks any form of transfer learning. Second, the advantage is not always there; in a few cases (SVM degree~3) performance degrades; a reason for such behavior is that overly complex models capture patterns that belong exclusively to the training set and may incur in a large number of misclassifications in a validation set. Third, SA exhibits poor performance, and does not provide any advantage during transfer learning. This is a phenomenon known as \textit{negative transfer} where the use of transfer learning leads to a loss of performance. The poor performance of SA can be explained as a result of the preliminary step to unify source and target into the same feature space; the common space does not achieve a sufficient overlap between the two distributions.

\begin{wrapfigure}{l}{0.5\textwidth}
\vspace*{-4mm}
\hspace*{0mm}
\includegraphics[scale=0.22]{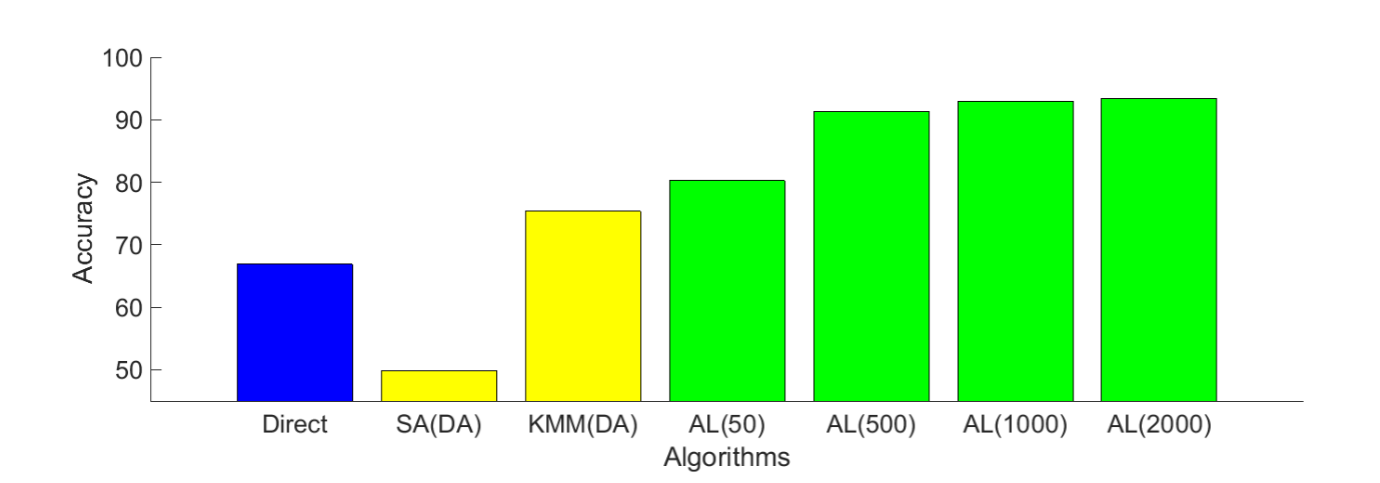}
\vspace*{-5mm}
\caption{\small Accuracy on target data using transfer learning and active learning. The last 4 columns show performance when active learning is invoked (with increasing budget).}
\label{fig:Supernova_DA_with_AL}
\vspace*{-4mm}
\end{wrapfigure}

An additional observation is that further enhancement of the target model can be achieved by combining transfer learning with an area of study in machine learning known as \textit{active-learning} \cite{Balcan09}. Active learning points to those few instances on the target set where knowing the class label with confidence suffices to attain an accurate model. Here, the system is allowed to ask the expert for the right class label on the selected event or instance (e.g., by running a spectroscopic analysis on the selected Supernova). The process works as follows. It begins by using the model $f_t(\mathbf{x})$ created by the transfer learning algorithm as the initial model for active learning. This is followed by an iterative process that queries the next most informative instance from the target dataset, and builds a new model after the last queried instance is added to the sample. The algorithm stops when it reaches a maximum cost i.e., when it runs out of budget. Figure~\ref{fig:Supernova_DA_with_AL} shows results for the automated classification of Supernovae Ia when transfer learning and active learning are combined synergistically. Results show a significant advantage gained when transfer learning is combined with active learning. Even with a modest budget size, the combination yields better performance than the use of transfer learning alone. Of course, growing the budget size provides increased performance gain.


\section{Summary and Conclusions}
\label{sc:conclusions}

This paper shows the value behind an emerging area of study in machine learning known as \textit{transfer learning}. The main idea is to leverage experience from previous learning tasks by transferring knowledge from a source domain to a target domain (Section~\ref{sc:introduction}). There are multiple approaches to transfer learning (Section~\ref{sc:types-tl}); examples include homogenous and heterogenous transfer; transfer mechanisms based on the differences between marginal and/or class-posterior distributions;  and finally, transfer learning based on the nature of the element being transferred.

The experimental study described in Section~\ref{sc:application-tl} shows an application of transfer learning on the identification of Supernovae~Ia (SNe Ia). The study shows that while transfer learning can yield substantial gains in performance, the technology itself is not always guaranteed to succeed. In the particular study described here, kernel mean matching shows an increase in accuracy performance when identifying SNe Ia, but subspace alignment leads to a loss in accuracy; for the latter case I mentioned that some approaches to transfer learning can lead to what is known as \textit{negative transfer}, a scenario where performance degradation is observed. The section also mentions the benefit of combining transfer learning with active learning as a means to add relevant labeled examples from the target dataset.


\vspace*{-2mm}
\subsection{Acknowledgments}
This work was partly supported by the Center for Advanced Computing and Data Systems (CACDS), and
by the Texas Institute for Measurement, Evaluation, and Statistics (TIMES) at the University of
Houston.



\vspace*{-2mm}
\section*{References}

\vspace*{0mm}

\begin{thebibliography}{9}

\bibitem{Balcan09}
Balcan, M.-F.; Beygelzimer, A.; Langford, J.
\newblock 2009.
\newblock Agnostic Active Learning.
\newblock Journal of Computer and System Sciences, 75(1):78 -- 89.

\bibitem{BenDavid10}
Ben-David, S.; Blitzer, J.; Crammer, K.; Kulesza, A.; Pereira, F.; Vaughan, J.
\newblock 2010.
\newblock  A Theory of Learning from Different Domains.
\newblock Machine Learning, vol. 79, pp. 151–175, 2010.

\bibitem{Blitzer08}
Blitzer, J.; Crammer, K.; Kulesza, A.; Pereira, F.; Wortman, J.
\newblock 2008.
\newblock  Learning Bounds for Domain Adaptation.
\newblock  Advances in Neural Information Processing Systems 20.
\newblock  Platt, J. C.; Koller, D.; Singer, Y.; Roweis, S. T.; Eds. Curran Associates, Inc., 2008, pp. 129–136.

\bibitem{Fernando14}
Fernando, B.; Habrard, A.;p Sebban, M.; Tuytelaars, T.
\newblock 2014.
\newblock Subspace Alignment for Domain Adaptation.
\newblock CoRR, vol. abs/1409.5241.

\bibitem{Goodfellow16}
Goodfellow, I; Bengio, Y; Courville, A.
\newblock 2016.
\newblock Deep Learning.
\newblock MIT Press.

\bibitem{Gretton09}
Gretton, A.; Smola, A.; Huang, J.; Schmittfull, M.; Borgwardt, K.; Sch¨olkopf, B.
\newblock  2009.
\newblock  Covariate Shift by Kernel Mean Matching.
\newblock  Dataset Shift in Machine Learning, vol. 3, no. 4.

\bibitem{Gupta16}
Dhar Gupta, K.; Pampana, R.; Vilalta, R.; Ishida, E. E. O.; de Souza, R. S.
\newblock 2016.
\newblock Automated Supernova Ia Classification Using Adaptive Learning Techniques.
\newblock IEEE Symposium on Computational Intelligence and Data Mining (CIDM-16), Athens, Greece.

\bibitem{Hastie09}
Hastie, T.; Tibshirani, R.; Friedman, J.
\newblock 2009.
\newblock The Elements of Statistical Learning: Data Mining, Inference and Prediction.
\newblock Springer, 2nd edition.

\bibitem{Li12}
Li, F.; Pan, S.J.; Jin, O.; Yang, Q.; Zhu, X.
\newblock 2012
\newblock Cross-Domain Co-Extraction of Sentiment and Topic Lexicons.
\newblock Proceedings of the 50th Annual Meeting of the Association for Computational Linguistics Long Papers, vol. 1. 2012. p. 410–19.

\bibitem{Pan10}
Pan, S.~J.; Yang, Q.
\newblock 2010.
\newblock A Survey on Transfer Learning.
\newblock IEEE Transactions on Knowledge and Data Engineering 22(10):1345--1359.

\bibitem{Quinonero09}
Quionero-Candela, J.; Sugiyama, M.; Schwaighofer, A.; Lawrence, N. D.
\newblock 2009.
\newblock Dataset Shift in Machine Learning.
\newblock The MIT Press.

\bibitem{Riess98}
Riess, A. G., et al.
\newblock 1998.
\newblock Observational Evidence from Supernovae for an Accelerating Universe and a Cosmological Constant.
\newblock The Astronomical Journal, AJ, V. 116, No. 3, 1009.

\bibitem{Sasdelli16}
Sasdelli, M., et al.
\newblock 2016.
\newblock Exploring the Spectroscopic Diversity of Type Ia Supernovae with DRACULA: A Machine Learning Approach.
\newblock Monthly Notices of the Royal Astronomical Society (MNRAS), 461, no.2, pp. 2044-2059, Oxford University Press.

\bibitem{Scholkopf97}
Sch¨olkopf, B.; Smola, A.; Muller, K.-R.
\newblock 1997.
\newblock Kernel Principal Component Analysis.
\newblock Berlin, Heidelberg: Springer Berlin Heidelberg, 1997, pp. 583–588.

\bibitem{Shimodaira00}
Shimodaira, H.
\newblock 2000.
\newblock Improving Predictive Inference Under Covariate Shift by Weighting the Log-Likelihood Function.
\newblock Journal of Statistical Planning and Inference, vol. 90, no. 2, pp. 227–244.

\bibitem{Vilalta13}
Vilalta, R.; Dhar Gupta K.; Macri, L.
\newblock 2013.
\newblock A Machine Learning Approach to Cepheid Variable Star Classification Using Data Alignment and Maximum Likelihood.
\newblock Astronomy and Computing Journal, Vol. 2, pp. 46-53. Elsevier. DOI: 10.1016/j.ascom.2013.07.002.

\bibitem{Weiss16}
Weiss, K.; Khoshgoftar, T.; Wang, D.
\newblock 2016.
\newblock A Survey of Transfer Learning.
\newblock Journal of Big Data 3:9 DOI 10.1186/s40537-016-0043-6.

\end{thebibliography}
\end{document}